\documentclass[11pt]{article}

	\usepackage{a4,geometry}

\usepackage{graphicx}
\usepackage{amsmath,amssymb,amsthm,mathtools}
\usepackage{paralist}
\usepackage{bm}
\usepackage{xspace}
\usepackage{url}
\usepackage{fullpage, prettyref}
\usepackage{boxedminipage}
\usepackage{wrapfig}
\usepackage{ifthen}
\usepackage{color}
\usepackage{xcolor}
\usepackage{framed}
\usepackage{algorithmic,algorithm}
\usepackage[pagebackref,colorlinks=true,pdfpagemode=UseNone,urlcolor=blue,linkcolor=blue,citecolor=violet,pdfstartview=FitH]{hyperref}
\usepackage{fullpage}

\usepackage{thmtools}
\usepackage{thm-restate}

\newtheorem{theorem}{Theorem}[section]

\newtheorem{lemma}[theorem]{Lemma}
\newtheorem{claim}[theorem]{Claim}
\newtheorem{corollary}[theorem]{Corollary}
\newtheorem{definition}[theorem]{Definition}
\newtheorem{proposition}[theorem]{Proposition}
\newtheorem{fact}[theorem]{Fact}
\newtheorem{example}[theorem]{Example}
\newtheorem{assumption}[theorem]{Assumption}
\newtheorem{observation}[theorem]{Observation}

\newcommand{\ignore}[1]{}

\newcommand{\R}{\mathbb R}

\newcommand{\eps}{\varepsilon}

\newcommand{\poly}{\mathrm{poly}}

\newcommand{\Exp}{\EX}
\newcommand{\floor}[1]{\lfloor#1\rfloor}

\newcommand{\Sec}[1]{\hyperref[sec:#1]{\S\ref*{sec:#1}}} 
\newcommand{\Eqn}[1]{\hyperref[eq:#1]{(\ref*{eq:#1})}} 
\newcommand{\Fig}[1]{\hyperref[fig:#1]{Fig.\,\ref*{fig:#1}}} 
\newcommand{\Tab}[1]{\hyperref[tab:#1]{Tab.\,\ref*{tab:#1}}} 
\newcommand{\Thm}[1]{\hyperref[thm:#1]{Theorem\,\ref*{thm:#1}}} 
\newcommand{\Fact}[1]{\hyperref[fact:#1]{Fact\,\ref*{fact:#1}}} 
\newcommand{\Lem}[1]{\hyperref[lem:#1]{Lemma\,\ref*{lem:#1}}} 
\newcommand{\Prop}[1]{\hyperref[prop:#1]{Prop.~\ref*{prop:#1}}} 
\newcommand{\Cor}[1]{\hyperref[cor:#1]{Corollary~\ref*{cor:#1}}} 
\newcommand{\Conj}[1]{\hyperref[conj:#1]{Conjecture~\ref*{conj:#1}}} 
\newcommand{\Def}[1]{\hyperref[def:#1]{Definition~\ref*{def:#1}}} 
\newcommand{\Alg}[1]{\hyperref[alg:#1]{Alg.~\ref*{alg:#1}}} 
\newcommand{\Ex}[1]{\hyperref[ex:#1]{Ex.~\ref*{ex:#1}}} 
\newcommand{\Clm}[1]{\hyperref[clm:#1]{Claim~\ref*{clm:#1}}} 

\renewcommand{\R}{\mathbb{R}}
\newcommand{\C}{\mathcal{S}}

\newcommand{\Var}{\mbox{Var}}

\renewcommand{\ignore}[1]{}

\def\eps{\varepsilon}
\def\epsilon{\eps}
\def\bar{\overline}
\def\floor#1{\lfloor {#1} \rfloor}

\def\Pr{\mathbf{P}}
\def\Exp{\mathbf{E}}
\def\Var{\mathbf{V }}

\def\chi{X}

\def\dist{\sf{dist}}

\newcommand{\specificthanks}[1]{\@fnsymbol{#1}}
\title{Detecting Character Dependencies in \\
 Stochastic Models of Evolution}
\author{Deeparnab Chakrabarty\thanks{Microsoft Research, {\tt dechakr@microsoft.com}}
\and Sampath Kannan\thanks{Department of Computer and Information Science, University of Pennsylvania {\tt kannan,ktian@upenn.edu}}
\and Kevin Tian\footnotemark[2]}
\date{\today}

\begin{document}
\maketitle

\begin{abstract}
Stochastic models of evolution (Markov random fields on trivalent trees) generally assume that different characters (different runs of the stochastic process) are independent and identically distributed. In this paper we take the first steps towards dealing with dependent characters.  Specifically, we consider various stochastic models of evolution ranging from the common ones used by biologists (such as Cavender-Farris-Neyman and Jukes-Cantor models) to very general ones where evolution of different characters and on different edges can be governed by different transition matrices. We also consider several models of dependence between two characters. In the most specific model, on each edge of the phylogeny the joint distribution of the dependent characters undergoes a perturbation of a fixed magnitude, in a fixed direction from what it would be if the characters were evolving independently. More general dependence models don't require such a strong `signal'. Instead they only require that on each edge, the perturbation of the joint distribution has a significant component in a specific direction. Our main results are nearly tight bounds on the induced or operator norm of the transition matrices that would allow us to detect dependence efficiently for most combinations of evolution and dependence model.   We make essential use of 
a new concentration result for multistate random variables of a Markov random field on arbitrary trivalent trees: we show that the random variable counting the number of leaves in any particular state has variance that is subquadratic in the number of leaves.

\end{abstract}

\thispagestyle{empty}
\newpage
\setcounter{page}{1}

\section{Introduction}\label{sec:intro}
Reconstructing the phylogeny or evolutionary tree of a set of organisms is a very important problem in biology~\cite{SS03,Fe04}. The general formulation of the problem is the following:  data corresponding to the species alive today is observed at the leaves of a certain unknown tree which is supposed to model the evolutionary progress thus far.  The goal is to find the best tree `fitting the data' under a specified objective function. Nowadays the most common type of data we observe is biomolecular sequences, i.e., DNA or protein sequences. Let 
 $\sigma_i$ be the sequence obtained from the $i^{th}$ species. Furthermore, these input sequences can be {\em aligned}, i.e., lined up in columns such that  for all $i,\, i',$ and for any position, or {\em character} as is called in the literature and what we will use henceforth, $j$, $\sigma_i[j]$   and $\sigma_{i'}[j]$ have a common evolutionary origin\footnote{This might sound circular to non-experts since alignment seems to require knowledge of the evolutionary process, but biologists have realized this process so successfully that it has become a standard technique in building phylogenies.}, where $\sigma_i[j]$ represents the $j^{th}$ symbol in $\sigma_i$

The most principled 
method of finding a phylogeny is to view the evolution of each position of the aligned DNA sequences as a stochastic process, more specifically, as a tree Markov random field whose parameters are chosen from a rich family of possible parameters.
Using the Maximum Likelihood objective function, the goal is to reconstruct the tree and most likely values of the parameters given the observed data at the leaves\cite{Fe81,HuCr97}. Under standard stochastic models of the evolutionary process 
{\em and} reasonable technical constrains on the transition matrices, 
considerable work \cite{FarachKannan, Erdos1, Erdos2, MosselRoch, Chang, Mo04, DaMoRo06} has been done to determine the number of characters needed to infer the phylogeny. {\em All of these works assume that the stochastic processes governing each character are  independent and identically distributed}.

The independence assumption across characters is too strong.
Dependence between characters arises because changes at one position of a DNA sequence or amino acid sequence are likely to be correlated with changes at other positions because of constraints on size, charge, hydrophobicity, etc. of the molecules involved \cite{MJH+13,MPL+11}. However, thus far to our knowledge, phylogeny reconstruction has not been studied under dependent characters. In fact,
{\em even the question of whether or not two given characters are independent is not understood.} 
Maddison~\cite{Ma90} and the references cited therein outline simple heuristic procedures, but no general procedures with provable properties have been proposed for detecting dependence, and indeed dependence has not been mathematically modeled so far in the literature.  Our paper makes a first step in addressing this. 

A tree Markov random field consists of an underlying rooted tree  $T$. A {\em character} (position in the biomolecular sequence) on such a tree is a stochastic process that takes on a value at each node from a set of finitely many {\em states} (genes {A,C,G,T}, or amino acids). At the root of $T$ the value is chosen from some initial distribution over the states.  Each parent passes on its state to its children. However, the value is `mutated' along each edge with probabilities given by a Markov transition matrix corresponding to the edge. For each character we  observe its state at each leaf of the tree. The question we consider in this paper is -- {\em given two such characters, are they independent?} Tree Markov random fields are a standard way to model the process of evolution, and 
all commonly-studied families of stochastic models are special cases of tree Markov random fields. Among the simplest are two-state, symmetric models, called the Cavender-Farris-Neyman (CFN)\cite{Ne71,Fa73,Ca78} models, where on any edge $e$, all characters have a symmetric $2\times 2$ transition matrix $M_{e}$. The Jukes-Cantor model is a $4$-state model where for any transition matrix there is a parameter $\epsilon$ that is the probability of any change of state \cite{JC69}. In this paper, we will look at a range of models of evolution to address the question above;
even the most restrictive model of evolution we consider in this paper is a generalization of all these standard biological models. 

 Our paper introduces some simple models of dependence among characters that seem well-suited to the biological application. The definition of these models themselves is one of the main contributions of this paper. If two characters are independent, then on each edge of the tree the matrix governing their joint evolution is just the tensor product of the marginal matrices. Two characters are dependent if this does not hold, i.e., that there are edges where the transition matrix for the joint character differs  from the tensor product of the marginals. But such a general kind of dependence might not even be detectable at the leaves since the dependence on one edge could be `canceled out' by the dependence on another edge. Thus we need to make an assumption that also appears to be biologically meaningful. We assume there is a `consistency' in the dependence between the characters on all edges. By this we mean that each row of the actual transition matrix governing the joint evolution of two dependent characters differs from the corresponding row of the tensor product matrix by a vector $v$ that is (roughly) in the same direction across all rows and across all edges. The detailed definition is given in the sequel. Biologically this makes sense because we expect that if there is dependence between two characters, then a certain subset of the joint states of the two characters should be consistently preferred across all edges of the tree to what would be with independent evolution. Put another way, suppose two characters have an affinity to be in the same state in a certain species, then it is fair to assume the affinity is present in its ancestors as well.  
 
A technical contribution of the paper is a concentration bound for tree Markov random fields. Let $Z$ be the random variable counting the number of occurrences of a character in a particular state at the leaves of a rooted tree. We show that as long as a certain natural norm of the transition matrices are bounded away from $1$ (by an arbitrary small constant amount), the variance of $Z$ is sub-quadratic in the number of leaves, and the expectation of $Z$ is linear in the number of leaves.  Tje technical challenge in proving this is to overcome the confounding dependence between the states of nearby leaves.
The other major technical challenge is to show that the `dependency signal' which occurs at every edge persists at the leaves even if it is subjected to different transformations at each edge of a root-leaf path.
In fact, as we show, this can't occur in general and we give bounds on the norm for which such a persistence does occur.



\section{Preliminaries and Statement of Results}\label{sec:prelims}
\noindent
{\bf Stochastic Model of Evolution.}
Let $T$ rooted at $r$ denote the underlying tree in a tree Markov random field. 
With little loss of generality we assume that the root has degree $2$ and every other internal node has degree $3$. 
A {\em character} maps the nodes of the tree to a set $\C$ of $s$ {\em states}  (for example, $\{0,1\}$, $\{A,C,G,T\}$, $\{\textrm{20 amino acids}\}$, etc.  A single character evolves `down' the tree as follows. At the root $r$ it has some distribution over its states, which need not be uniform. However, it is sufficient to consider the initial distribution to be uniform due to the mixing properties of the stochastic process. With every edge $e = (u,v)$ of $T$, is associated a stochastic $s\times s$  transition matrix $M_e$ that  governs the evolution of the character. More precisely,  $\Pr[X_v = b| X_u = a] = M_e(a,b)$. 

Specific biological models assume that these matrices are drawn from special types of stochastic matrices. For instance, the Cavender-Farris-Neyman (CFN)~\cite{Ne71,Fa73,Ca78} model for binary states ($s=2$) assumes that on each edge all characters have the same symmetric transition matrices. Thus a single scalar (the probability of mutation) determines the transition matrix on any edge; this scalar is usually a  measure of the time duration represented by the edge. The Jukes-Cantor~\cite{JC69} model is a simple generalization to 4-state characters and the  Kimura~\cite{Kimura80} model is determined by 2 parameters rather than 1. In our work, we consider models of evolution at 3 levels of generality, listed below. All the above biological models lie in the most restrictive level. One reason we consider the more general models is because they lead to mathematically interesting problems whose solutions might be applicable in other contexts beyond phylogenies. Note that in this work, we distinguish between an independent case and a dependent case; the required properties listed below apply to the transitions in the independent case. The transitions in the dependent case differ from transitions which satisfy the listed properties by an `error matrix' which we specify below.

\begin{asparaenum}[noitemsep,nolistsep]
\item[\bf Shared Eigenbasis.] Our most restrictive model assumes that all transition matrices are positive semi-definite (PSD) and have the same eigenbasis on every edge; this is true for all biological models studied so far.
\item[\bf PSD.] At a greater level of generality,  we do not require the PSD matrices for a character to have the same eigenbases on all edges.
\item[\bf Doubly stochastic.] In this model we just assume all transition matrices are doubly stochastic.
\end{asparaenum}\smallskip

\noindent
The parameter that governs our results is the following $1\to 1$ norm 
of transition matrices:
$||M|| := \sup_{0\neq x\bot \mathbf{1}}||x^\top M||_1/||x||_1$. 
{\em We assume $||M||\leq \lambda < 1$ for some constant $\lambda$.}
It is easy\footnote{$v$ be an eigenvector corresponding to eigenvalue $\lambda \!< \!1$. $M$ is stochastic, so $v\bot \mathbf{1}$ and $Mv = \lambda v$ implies $||M||\geq |\lambda|$.} to see that $||M||$ is always at least the second eigenvalue (in absolute value) of $M$; our above assumption implies
$\lambda_2(M)\leq \lambda$ as well.
In order to detect dependence we will need increasingly tighter upper bounds on $\lambda_2(M)$ as we move to more general models of evolution. \smallskip

\noindent
{\bf The Dependence Model.}
Let $X$ and $Y$ be two characters and let $X_u$ and $Y_u$ denote the states of these characters at node $u$ of $T$.  If $X$ and $Y$ evolve independently, then the transition matrix governing the evolution of the joint variable $(X,Y)$ across edge $e$ of the tree is given by the matrix $M_e \otimes N_e$ where $M_e$ and $N_e$ are the $s\times s$ transition matrices associated with the individual characters. Note that we allow different characters to have different transition matrices.  If $X$ and $Y$ are not independent, then we assume the following dependence model.
Firstly, we assume that the joint random variable $(X,Y)$ evolves via a Markovian process. This is standard in biology where mutation is assumed to be history independent.
So, for every edge $e$ there exists an $s^2 \times s^2$ transition matrix $P_e$ such that
$\Pr[(X_v,Y_v) = (a',b') | (X_u,Y_u) = (a,b)] = P_e((a,b),(a',b'))$. Furthermore, we assume a {\em consistent preferred direction} dependence model where the joint evolution of the two characters tends to bias probabilities in a preferred direction in comparison to the situation when they evolve independently. We model this by making assumptions on the `deviation' matrix $D_e := P_e - M_e \otimes N_e$. In the simplest, but already non-trivial case that we call the {\bf uniform rank-1 dependence} model,  we assume that the deviation matrix $D_e = D$ for all edges, and furthermore, $D = \mathbf{1}d^\top$ for some $s^2$-dimensional vector $d$ with $||d||_1 \geq \delta > 0$ for some known  parameter $\delta$. This stringently models  situations where there are preferred states and every transition biases the distribution by the same vector in favor of the preferred states regardless of the starting state or edge. 
We also investigate a generalization of the uniform rank-1 dependence that we call the {\bf directional-drift dependence} model.   Here we assume there exist some direction $d^*$ such that every row of every deviation matrix $D_e$ has an inner product of at least $\delta$ with $d^*$. In addition, the norm of any row of any of these matrices is at most a constant.
For ease of presentation we use the uniform rank-$1$ model almost throughout the paper, only discussing the more general model in the last section.\footnote{Throughout the paper, we are concerned with detecting dependence between a pair of characters. However, it is not hard to generalize our models and results to a constant subset of characters
where instead of pairs $\C\times \C$, we would be dealing with random variables over a larger domain. For simplicity, we just stick with pairs.}  

\noindent
{\bf Informal Statement of Results for uniform rank-1 dependence:}
\begin{enumerate}
\item In the shared eigenbasis model, we can detect dependence with no further assumptions.
As stated above, this includes all the major models of evolution studied so far.

\item In the PSD model, if all single-character transition matrices have $\lambda_2 < 0.797$, then we can detect dependence. However, there exists examples of trees with PSD transition matrices with $\lambda_2 \geq 0.832$ and yet the distribution on the leaves is indistinguishable from the case of independent evolution.

\item In the doubly-stochastic model, we can detect dependence if all transition matrices have $\lambda_2 \leq \frac12$. We cannot prove `better ' negative results  than for the PSD case.
\end{enumerate}

\smallskip

\noindent
{\bf Informal Statement of Result for directional-drift dependence:}

\noindent
If each row of $D_e$ has length at most $\delta/\beta$, then we can allow $\lambda_2 \leq \frac{\beta}{\frac12 + \beta}$.

\ignore{Throughout the paper, we will be concerned with detecting dependence between a pair of characters. However, it is not hard to generalize our models and results to a constant subset of characters
where instead of pairs $\C\times \C$, we will be dealing with random variables over a larger domain. For simplicity, we just stick with pairs.
}

\def\S{\mathcal P}

\section{The Tester, Analysis Roadmap, and Technical Challenges}
The input to our dependency testers are the values of the characters at the leaves of the phylogeny. 
Our tester is extremely simple: For each ordered pair of states, we count the number of leaves that have that pair. If there is a `large discrepancy' in this number, the characters are dependent.

\medskip

\fbox{\begin{minipage}{.9\textwidth}
\noindent
{\em Algorithm} ~{\small \bf Dependence detection}\\
\noindent
\underline{Input:} States of the characters at $n$ leaves\\
\noindent
\underline{Parameter:} Precision parameter $\eps$

\medskip
\noindent
1. For $i \in \C\times\C$, let $Z_i$ denote the number of leaves with $(X_l,Y_l) = i$. \\
2. If for any $i,j \in \C\times\C$, $|Z_i - Z_j| \geq \eps n$, output that the characters are dependent. Else, output that the characters are independent.
\end{minipage}} 
\medskip

\noindent
Note that we have used a single index $i$ to denote a pair of states.
We now briefly outline the analysis and the challenges involved.

We prove a concentration bound for the overall distribution of state pairs at the leaves. For all ordered pairs $i$, we need that the number of leaves $Z_i$ that have the state $i$, is concentrated around its mean. This is not trivial since $Z_i$ is a sum of indicator random variables that are not independent, because in a tree Markov random field, even the state of one character at `near by' leaves are highly correlated. We obtain concentration by upper-bounding the second moment (\Thm{main}) which is done via a coarse but sufficiently good upper bound on the variance (\Lem{bndvar}).
We also show (\Lem{independent}) that when the characters are independent, we {\em expect} each joint state to be almost equally likely.
Since the norms of the transition matrices are bounded away from $1$,  we expect rapid mixing and the leaf state to be close to stationary distribution, which by the assumption of double-stochasticity is uniform. 

In the case of dependent characters, we show that a large discrepancy (\Lem{dependent}) indeed occurs (in expectation) for at least one pair of states.  This is nontrivial 
since different edges have different transition matrices, and the effect of one matrix's deviation may cancel the effect of its predecessors. Indeed, in the PSD model, we show that this can happen even when $\lambda_2 \geq 0.832$ (\Lem{models:negative}). 
However, if all matrices share the same eigenbasis, then such a `bad case' cannot occur (\Lem{models:shared}), and so the shared eigenbase model doesn't need any further assumptions.  In the doubly stochastic model, an upper bound of $0.5$ suffices (\Lem{models:doubly}) to detect dependency.
 For the PSD model, an upper bound $\lambda_2 \leq 0.797$ suffices, and this is more subtle to show. 
 To do so, we prove a lower bound on a quantity $v^\top Av$ where $A$ is a product of $k$ PSD matrices (and therefore, not necessarily PSD) and $v$ is a vector perpendicular to the all ones vector. We show (\Lem{models:psd:pessimistic}) that this quantity is at least $-(\lambda\cos(\pi/k+1))^k||v||^2_2$; this result may be of independent interest. 
 We leave open the question of finding the exact value in $[0.797, 0.832]$ at which dependence can be detected in the PSD model.
Finally, in \Thm{directional}, we show that in the directional-dependence model, we can detect dependence when $\lambda^*$ is bounded by a function of $\delta,\beta$. 

\section{Bounding the Variance}\label{sec:var}

Fix an $i\in \S = \C\times \C$.
Let $Z$ be the random variable counting the number of leaves of $T$ in state $i$.
For a random variable $X$, let $\Exp[X]$ denote its expectation and $\Var(X)$ its variance.
Recall $\lambda < 1$ is an upper bound on the norm of any of the transition matrices $P_e$ on the edges $e$.
In this section, we prove the following theorem.
\begin{theorem}\label{thm:main}
Given any $n$ leaf trivalent tree $T$,  $\Var(Z) = O(n^{2 - 2\log_2(1/\lambda)})$.
\end{theorem}

\noindent
For any vertex $v$ in $T$, let $Z_v$ to be the number of leaves in the sub-tree of $T$ rooted at $v$ in state in $i$; so $Z = Z_{\rm root}$. Let $L_v$ denote the leaves in the subtree rooted at $v$. For a leaf $\ell \in L_v$, let $\dist(v,\ell)$ denote the number of edges on the path from $v$ to $\ell$ in the tree.
Define
\begin{equation}
\label{eq:Lambda}
\Lambda(v) \triangleq 2\sum_{\ell \in L_v} \lambda^{\dist(v,\ell)}
\end{equation}
The following claim bounds $\Lambda(v)$ at any vertex; the proof can be found in Appendix \ref{sec:omitted}
\begin{restatable}{claim}{bounddelta}
\label{claim:bounddelta}
For any vertex $u$ with $n$ leaves in its subtree, $\Lambda(u) \leq O(n^{1-\eta})$ where
$\eta = \log_2(1/\lambda)$.
\end{restatable}

\noindent
The following lemma bounds the variance in terms of the $\Lambda$'s. We first use the lemma to prove the theorem and then go on to prove the lemma.
\begin{lemma}
\label{lem:bndvar}
$\Var(Z) \leq \frac{1}{2}\sum_{v\in V(T)\setminus{\rm root}} (\Lambda(v))^2$.
\end{lemma}
\def\V{\mathcal{V}}
\noindent
\begin{proof}[{\bf Proof of Theorem \ref{thm:main}}]
Let $\V(n)$ be a function that denotes the maximum value of $\sum_{v\in V(T)\setminus r} \Lambda^2(v)$ 
over all $n$-leaf binary trees. By \Lem{bndvar}, we want a subquadratic upperbound on $\V(n)$.
Let $u$ be the {\em centroid} of $T$. That is, $n/3 \le |L_u| \le 2n/3$.
It is easy to see this is well defined. Let $T_u$ denote the subtree of $T$ rooted at $u$, and let 
$T'_u$ denote the subtree of $T$ with all descendants of $u$ deleted. Note that both $T_u$ and $T'_u$ are binary trees, and have $\rho n$ and $(1-\rho)n$ leaves for $\rho \in [1/3,2/3]$. 
By definition, 
$\sum_{v\in V(T_u)\setminus u} \Lambda^2(v) \le \V(\rho n)$ and $\sum_{v\in V(T'_u)\setminus r} \Lambda^2(v) \le \V((1-\rho) n)$.

Suppose $u = u_0,u_1,\ldots,u_r = r$ is the unique path from $u$ to $r$ in $T$.
Note that the $\Lambda(v)$'s in tree $T'_u$ are the same as in tree $T$ for all vertices except the $u_i$'s.
For each $u_i$, $\Lambda(u_i)$ in the tree $T$ is that in $T'_u$ plus $\lambda^i\cdot \Lambda(u)$
Thus, we have

\begin{align*}
\sum_{v\in V(T)\setminus r} \Lambda^2(v) & \le & \V(\rho n) +  \V((1-\rho) n) ~+~ \sum_{i=0}^r \Big( (\Lambda(u_i) + 2\lambda^i\Lambda(u))^2 - \Lambda^2(u_i)\Big) \\
& = & \V(\rho n) +  \V((1-\rho) n) ~+~ 4\Lambda(u)\sum_{i=0}^r \lambda^i\Lambda(u_i) ~ + ~ 4\Lambda^2(u)\sum_{i=0}^r \lambda^{2i}
\end{align*}
From Claim \ref{claim:bounddelta}, we can bound $\Lambda(u_i)$ by $O(n^{1-\eta})$ for $i=0,...,r$. So we get the following recurrence for $\V(n)$
$\V(n) \le \V(\rho n) + \V((1-\rho) n) + O(n^{2 - 2\eta})$
\noindent
which evaluates to $\V(n) = O(n^{2 - 2\eta})$. 
\end{proof}

\paragraph{Proof of \Lem{bndvar}}
For a vertex $v$ and two states $j,k \in \S$, define
\begin{equation}
\label{eq:deltajk}
\Delta_v(j,k)  \triangleq |\Exp[Z_v| X_v = j] - \Exp[Z_v| X_v = k]|
\end{equation}
\noindent
The following claim relates $\Delta_v$ with $\Lambda(v)$. 
\begin{claim}
\label{clm:deltalessthanlambda}
For any vertex $v$, and for any two states $j,k\in \S$, we have $\Delta_v(j,k) \leq \Lambda(v)$.
\end{claim}
\begin{proof}
Fix a vertex $v$ and a leaf $\ell\in L_v$. Let $e_1,e_2,\ldots,e_{\dist(v,\ell)}$ be the edges on the path from $v$ to $\ell$. 
Let $P$ denote the matrix $P_{e_1}\cdot P_{e_2}\cdots P_{e_{\dist(v,\ell)}}$; this is the transition matrix from $v$ to leaf $\ell$. In particular, $P$ is row-stochastic (row entries add up to $1$).
We use the following simple fact about row stochastic matrices; a proof of this can be found in Lemma 4.12 in \cite{peres-book} or Appendix \ref{sec:fact}.
\begin{restatable}{fact}{lemfact}
\label{lem:fact}
For any two row stochastic matrices $P_1$ and $P_2$, we have $||P_1P_2|| \le ||P_1||\cdot||P_2||$.
\end{restatable}
\noindent
Fact \ref{lem:fact} implies that $||P||$ is at most $\lambda^{\dist(v,\ell)}$.  
In particular, this shows that for any vertex $v$, for any leaf $\ell\in L_v$ at a distance $\dist(v,\ell)$, and for any states $j,k\in \C$,  
we have
$|\Pr[X_\ell=i | X_v = j] - \Pr[X_\ell = i | X_v = k]| \le ||x^TP||_1 \le 2\lambda^{\dist(v,\ell)}$,
where $x$ is the vector with $x_j = 1$, $x_k = -1$, and $x_s = 0$ otherwise. 
The claim follows by noting
\begin{align*}
\Delta_v(j,k) ~ = &\quad  |\sum_{\ell\in L_v} \left(\Pr[ X_\ell = i | X_v = j] - \Pr[ X_\ell = i | X_v = k] \right) | \\ 
~ \le &\quad \sum_{\ell\in L_v} |\Pr[ X_\ell = i | X_v = j] - \Pr[ X_\ell = i| X_v = k] | \leq 2\sum_{\ell \in L_v}\lambda^{\dist(v,\ell)}
\end{align*}\end{proof}
\noindent
Now we can finish the proof of \Lem{bndvar}.
Fix any vertex $u$.
Recall that  $T_u$ denotes the subtree of $T$ rooted at $u$ and $Z_u$ is the number of leaves in $L_u$ in state $i$.
We now show using induction on the height of $T$ that for any state $j\in \S$, $\Var(Z_u|X_u = j) \le \frac{1}{2} \sum_{v\in V(T_u)\setminus u} \Lambda^2(v)$.
This proves the lemma with $u$ as the root, and summing over all the conditional events.

Note that the claim is vacuously true when $u$ is a leaf since both LHS and RHS are $0$.
Let $u$ have children $v_1,\ldots,v_q$ (if the tree is binary, $q=2$, but this lemma holds for any tree).
Assume we have proved the inductive claim for the $v_i$'s. Note that conditioned on $X_u$, the random variables $Z_{v_1},Z_{v_2},...$ are independent, since they count over leaves on disjoint subtrees. Therefore, for any $j\in \S$,
$\Var(Z_u|X_u = j) = \sum_{i=1}^q \Var(Z_{v_i}|X_u = j)$.

We now show that for any parent-child pair $e = (u,v_i)$ and any state $j\in \S$,  we have
\begin{align}
\Var(Z_{v_i}|X_u = j) = \sum_{k\in \S} P_{jk}\Var(Z_{v_i}|X_v = k) + \frac{1}{2}\sum_{k\neq k' \in \S} P_{jk}P_{jk'}\Delta^2_{v_i}(k,k')\label{eq:var}
\end{align}
\noindent
where $P_{jk} = \Pr[X_{v_i} = k| X_u = j] = P_e(j,k)$.  \eqref{eq:var} suffices to complete the proof. By induction, the first summand in the RHS is at most $\frac12\sum_{w\in T_{v_i}\setminus v_i}\Lambda^2(w)$. From Claim \ref{clm:deltalessthanlambda}, we have $\Delta_{v_i}(k,k') \leq \Lambda(v_i)$ and $\sum_{k\neq k'}P_{jk}P_{jk'} \leq (\sum_{k\in \S}P_{jk})^2 = 1$, thereby giving that the second summand in the RHS of \eqref{eq:var} is at most $\frac12\Lambda^2(v_i)$. Together, we get $\Var(Z_{v_i}|X_u = j) \leq \frac12\sum_{w\in T_{v_i}}\Lambda^2(w)$, and by adding over all $v_i$, $1\leq i\leq q$, we are done.
The equality \eqref{eq:var} follows via a straightforward calculation which can be found in Appendix \ref{sec:eqvar}.
This completes the proof of \Lem{bndvar}.

%


%

\section{Analysis of the Tester in Uniform Rank-1 Model}\label{sec:models}
Let $\mu_0$ be the state at the root. Let $\mu$ be the uniform distribution over the $s^2$ states.  Let $r_0 = \mu_0 - \mu$ be the {\em error vector} at the root. 
Recall $P_e$ is the transition matrix of the joint random variable $(X,Y)$ at edge $e$. We write $P_e = Q_e + D$ where $Q_e = M_e\otimes N_e$ and $D$ is
the zero-matrix if the characters are independent, and $D = {\mathbf 1}d^\top$ in case the characters are dependent.

Our goal in this section is to establish the following theorem\footnote{Note that $\lambda_2(Q_e) \leq \max(\lambda_2(M_e),\lambda_2(N_e))$}, asserting the correctness of the tester for the uniform rank-$1$-model.

\begin{theorem} \label{thm:positive}
	Under each of the following evolutionary models, under the listed assumptions on the norm on the transition matrices, Algorithm {\small \bf Dependence Detection} is correct with $1- 1/\poly(n)$ probability. 
	\begin{asparaenum}[noitemsep,nolistsep]
		\item[\bf Shared Eigenbasis.] No extra assumption on $Q_e$ is needed.
		\item[\bf PSD.] If $\lambda_2(Q_e) \leq 0.797$. 
		\item[\bf Doubly stochastic.] If $\lambda_2(Q_e) \leq 0.5$. 
	\end{asparaenum} 

\end{theorem}
For a leaf $\ell$, let $\mu_\ell$ be the distribution at the leaf, and $r_\ell = \mu_\ell - \mu$ be the error vector at the leaf. 
Let $(e_1,e_2,\ldots,e_{\dist(\ell)})$ be the path from the root to the leaf $\ell$.
Then, if the characters are independent, we get
\begin{equation}\label{eq:r:independent}
	r_\ell^\top = r_0^\top \left( \prod_{k=1}^{\dist(\ell)} Q_{e_k} \right)
\end{equation}
and if the characters are dependent, we get
\begin{equation}\label{eq:r:dependent:full}
			r_\ell^\top = d^\top + \sum_{i=1}^{\dist(\ell)} d^\top \left(\prod_{k=i+1}^{\dist(\ell)-1} Q_{e_k} \right) + r_0^\top \left( \prod_{k=1}^{\dist(\ell)} Q_{e_k} \right)
\end{equation}
		
We first prove a lemma to show that when the character pair evolves independently, the distribution of state pairs at the leaves is close to uniform.

\begin{lemma}\label{lem:independent}
If the characters are independent, then for all $i\in \S$, we have $|\Exp[Z_i] - n/s^2| \leq O(n^{1-\beta})$
for some constant $\beta$ depending on $\lambda$, the upper bound on the norms of all the transition matrices.
\end{lemma}
\begin{proof}
By our assumption,  $||Q_{e_i}||\leq \lambda$ for all $i$. Substituting in \eqref{eq:r:independent}, and using Fact \ref{lem:fact}, 
we get $||r_\ell||_1 \leq \lambda^{\dist(\ell)}||r_0||_1 \leq 2\lambda^{\dist(\ell)}$ since $||r_0||_1 \leq ||\mu||_1 + ||\mu_0||_1 = 2$. In turn, this implies $|\langle r_\ell,e_i\rangle| \leq 2\lambda^{\dist(\ell)}$. 
Note for any $i$, we have
$\Exp[Z_i] = \sum_{\ell} \langle \mu_\ell,e_i\rangle = n/s^2 + \sum_\ell\langle r_\ell,e_i\rangle \leq n/s^2 + \Lambda(\textrm{root})$, and the lemma follows from  Claim \ref{claim:bounddelta}.
\end{proof}
Next, we prove a contrasting lemma for the dependent case, depending on the model of evolution. In each case, we show that there is a deviation from the uniform in the distribution at the leaves.  In particular, we  exhibit $r^* \in \R^{s^2}$ whose coordinates sum up to zero with each entry in $[-1,+1]$ such that there is some $\eps > 0$ satisfying

\begin{equation}\label{eq:goal}\tag{{Deviation}}
\textrm{For all $\ell$, we have} \quad \langle r_\ell,r^*\rangle \geq \eps.
\end{equation}

We prove the following lemma, under the assumption that this equation is satisfied. In later subsections we demonstrate $r^*$ in every model of evolution.

\begin{lemma}\label{lem:dependent}
If under a model of evolution we obtain an $r^*$ and $\eps$ satisfying~\eqref{eq:goal}, then there exists $i,j\in \S$ such that 
$|\Exp[Z_i] - \Exp[Z_j]| \geq \eps n$ where $\eps$ is a constant depending on $\delta$ and $s$.
\begin{proof}
Let $\bar{\mu} := \frac{1}{n}\sum_{\ell}\mu_\ell$.
Observe that $\langle\bar{\mu},r^*\rangle \geq \eps$ as well. Since $r^*$ is a convex combination of vectors of the form $\{e_i-e_j\}$ where $e_i$ is the indicator vector for pair $i$, we get there exists $(i,j)$ such that $\langle \bar{\mu},(e_i - e_j)\rangle \geq \eps$. But $\langle\bar{\mu},e_i\rangle$ is precisely $\Exp[Z_i]/n$ since $\langle \mu_\ell,e_i\rangle$ indicates the probability leaf $\ell$ is in state $i$. Therefore \eqref{eq:goal} implies that there exists a pair $i$ and $j$ such that  $\Exp[Z_i] - \Exp[Z_j] \geq \eps n.$
\end{proof}
\end{lemma}
%

	\begin{proof}[Proof of Theorem \ref{thm:positive}]
		These follow from \Lem{independent} and \Lem{dependent} (with appropriate use of Equation \eqref{eq:goal}) using \Lem{models:shared}, \Lem{models:psd}, and \Lem{models:doubly} given below, \Thm{main} and Chebyshev's inequality.
	\end{proof}

\subsection{Shared eigenbasis model}\label{sec:models:shared}
Recall in this model we assume if characters are independent, then the transition matrices $M_e$ are PSD and share the same eigenbasis over all edges. This implies the matrix $Q_e = M_e\otimes N_e$ also is PSD and have the same eigenbasis across all edges. 
\begin{lemma} \label{lem:models:shared}
In the shared-eigenbasis model, for each leaf $\ell$, $\langle r_\ell,d\rangle \geq \|d\|^2(1-\lambda^{\dist(\ell)})$.
Thus in \Eqn{goal},  $r^* = d$ and $\eps = \|d\|_2^2(1-\lambda) \geq \delta^2(1-\lambda)/s$ suffices. 
\end{lemma}
\begin{proof}
We can multiply both sides of \Eqn{r:dependent:full} by $d$ to get 
$\langle r_\ell^\top d = d^\top d + \sum_{i=1}^{\dist(\ell)} d^\top A_i d + r_0^\top B d$, where
$A_i =\prod_{k=i+1}^{\dist(\ell)-1} Q_{e_k}$ while $B = \prod_{k=1}^{\dist(\ell)} Q_{e_k}$. The main observation is that if the $Q_e$'s share eigenbase, then products of these matrices are also PSD. Thus, each $A_i$ is PSD implying the second sum is $\geq 0$. The final term $|r^\top_0 Bd| \leq||r^\top_0||_\infty ||Bd||_1 \leq \lambda^{\dist(\ell)}$, by Cauchy-Schwartz, and the second inequality follows from since $||B||\leq \lambda^{\dist(\ell)}$. 
\end{proof}
%
%
%
%

%
%
%
\subsection{Positive semi-definite model}\label{sec:models:psd}
Recall that in this model each $M_e$ is PSD, and thus $Q_e$ is PSD as well. 
\begin{lemma} \label{lem:models:psd}
    In the PSD model, if $\lambda_2(Q_e) \leq \lambda^* < 0.797$, then 
for each leaf $\ell$, $\langle r_\ell,d\rangle \geq \epsilon(\lambda^*) > 0$.
\end{lemma}	
	\begin{proof}
		We expand \Eqn{r:dependent:full} to get
	(we ignore the last term since it vanishes with $\dist(\ell)$.)
		\begin{equation}\label{eq:r:innerproduct}
			\langle r_{\ell}, d \rangle = \|d\|_2^2 + \sum_{i=1}^{\dist(\ell)-1} d^\top \left( \prod_{k=i+1}^{\dist(\ell)} Q_{e_k} \right) d
		\end{equation}
		Note that each term in the sum is correlated, since they use the same matrices $Q_{e_k}$. To lower bound this product, we will relax this restriction, and allow that each term choose its own matrices.
		In particular, we will use the following lemma:
		
		\begin{lemma} \label{lem:models:psd:pessimistic}
			Suppose $A_1, \ldots, A_k$ are $k$ positive semi-definite transition matrices, with second eigenvalue bounded by $\lambda^*$, and let $v$ be a vector with entries summing to 0. Then
				\begin{equation}\label{eq:psd:pessimistic}
					v^\top (A_1 \cdots A_k) v \geq -(\lambda^*)^k \cos^{k+1} \left( \frac{\pi}{k+1} \right) \|v\|_2^2
				\end{equation}
			\end{lemma}
			\noindent
				Using this lemma, we can now bound
						$r_\ell^\top d \geq \|v\|_2^2 \left( 1 - \sum_{k=2}^\infty (\lambda^*)^k \cos^{k+1}\left( \frac{\pi}{k+1} \right) \right)$.
				Note that the paranthesized expression in the RHS can be lower bounded, for any integer $N\geq 2$, by $\left(1 - \sum_{k=2}^N  \cos^{k+1}\left( \frac{\pi}{k+1} \right)- \frac{(\lambda^*)^{N+1}}{1-\lambda^*} \right)$.
			    For instance if $N=2$, we get that if $\lambda^*\leq 2/3$, then the expression is lower bounded by $1/18$. Numerically, we obtained the best tradeoff at $N=8$ where $\lambda^* < 0.797$ implies
			    the expression is $> 0$.
			\end{proof}
			\noindent
			\begin{proof}[{\bf Proof of Lemma \ref{lem:models:psd:pessimistic}.}]
				We note that to minimize this, we are essentially looking to make $A_1 \cdots A_k v$ be a long vector pointing away from $v$. To do this, we can assume that each $A_i$ is a scaled projection onto a fixed vector $u_i$. Suppose some $A_i$ is not. Then let $u_i$ be a unit vector in the direction of $A_i \cdots A_k v$, and replace $A_i$ with a projection onto $u_i$ and a scaling by $\lambda^*$ without reducing the length of the resulting vector. Then we can let $\theta_i$ be the angle between $A_i \cdots A_k v$ and $A_{i+1} \cdots A_k v$, and $\theta_0$ the angle between $A_1 \cdots A_k v$ and $-v$. Then we have
					$|v^\top A_1 \cdots A_k v| = (\lambda^*)^k \left(\prod_{i=0}^k \cos(\theta_i)\right) \|v\|_2^2$.
				Finally, using the concavity and monotonicity of the cosine function in the domain $[0, \pi/2]$, and the fact that the total projections go from $v$ to $-v$, so $\sum \theta_i \geq \pi$, we conclude that to minimize this, each $\theta_i$ should be equal and so each $\theta_i = \frac{\pi}{k+1}$. 
			\end{proof}

Note that the value $0.797$ is not exact, even for this bound we have given, and better bounds may exist. 
However, we cannot allow $\lambda^*$ to be arbitrarily close to 1 which is encapsulated in the following lemma. 
We prove this in Appendix \ref{sec:negative}.
\begin{lemma}\label{lem:models:negative}
	In the PSD model it is not always possible to detect dependence at the leaves, even if $\lambda_2(Q_e) \leq 0.832$ for all $e$.
\end{lemma}
\subsection{Doubly-stochastic model}\label{sec:models:doubly}
In this model we simply assume the transition matrices are doubly stochastic. We show that if $\lambda_2(Q_e) < 1/2$, then we can detect dependence.
\begin{lemma} \label{lem:models:doubly}
In the doubly-stochastic model, each leaf $\ell$ has that $r_\ell$ satisfies 
	$\langle r_\ell,d \rangle \geq  \left( 1 - \frac{\lambda^*}{1 - \lambda^*}\right) \|d\|_2^2$.
	Thus \Eqn{goal} is satisfiable for constant $\eps > 0$  if $\lambda^* < 1/2$.
	\begin{proof}
		We will use a similar approach to \Lem{models:psd}. We will again use \eqref{eq:r:innerproduct}, and write
			\[ \langle r_\ell, d \rangle = \|d\|_2^2 + \sum_{i=1}^{\dist(\ell)-1} d^\top \left( \prod_{k=i+1}^{\dist(\ell)} Q_{e_k} \right) d \geq \|d\|_2^2 \left( 1 - \sum_{i=1}^\infty (\lambda^*)^i \right) = \|d\|_2^2 \left( 1 - \frac{\lambda^*}{1 - \lambda^*} \right)\]
		where we lower bounded
$d^\top Q_{e_{i+1}} \cdots Q_{e_{\dist(\ell)}} d$ by $- (\lambda^*)^{\dist(\ell) - i - 1} \|d\|_2^2$, since all eigenvalues in the space of error vectors are bounded in absolute value by $\lambda^*$. 
%
%
%
	
	\end{proof}
\end{lemma}

\section{Directional-drift Dependence Model}\label{sec:generalized}

Here, we generalize the error model, in the PSD evolution model, using the directional-drift dependence model which we now describe. We recall that PSD model generalizes the Shared Eigenbases model, which itself generalizes all stochastic models studied in the literature.
In this model, there is a fixed direction $d^*$, such that every row of each error matrix has the following properties: (1) $\|d^\top\| \leq \delta/\beta$, and (2) $\langle d, d^* \rangle \geq \delta$ for a significant $\delta$ and a constant $\beta$. 

\begin{theorem}\label{thm:directional}
	In the PSD evolutionary model with the directional-drift dependence model above, if all transition matrices have norm bounded by $\lambda^*(\beta) = \frac{\beta}{\frac12 + \beta}$, then {\small \bf Dependence detection} is correct.
\begin{proof}
	This theorem will again follow from Lemmas \ref{lem:dependent} and \ref{lem:independent}, through the use of Equation \eqref{eq:goal}, with $r^* = d^*$ and $\eps(\beta,\lambda)$.
\end{proof}
\end{theorem}

Let us first examine now what happens in one step, when we start with a vector $\vec{\mu} + \vec{r}$, and apply the transform $P_e = Q_e + D_e$,

	\[ (\mu + r)^\top \,\mapsto\, \mu^\top + r^\top Q_e + (\mu + r)^\top D_e \]

When $D_e = \mathbf{1} d^T$, we see the last term is precisely $d^\top$. Now, however, we get some vector $d_e$ which has $\| \vec{d} \|_2 \leq \delta/\beta$ and $\langle \vec{d}_e, \vec{d}^* \rangle \geq \delta$. We will use primarily the fact that this added vector has these properties. As before, we will view the transform in the error space, where the transform is

	\[ r^\top \,\mapsto\, r^\top Q_e + d_e \]

Our approach to show that the detection of dependence is possible here will be by induction. In particular, we will show that for each node other than the root, there is some $x^* = x^*(\beta)$ such that if the distribution at the node $v$ is $\mu + r_v$, then $\langle r_v, d^* \rangle \geq x^* \delta$. This is true for the direct children of the root, as the distribution is precisely $\mu + d_e$ where $e$ is the edge connecting to the root. By hypothesis, $\langle d_e, d^* \rangle \geq \delta$. This gives us the base case for induction.

Before we prove the general case, we will first observe that $\|r\|_2 \leq \frac{1}{1 - \lambda^*} \frac{\delta}{\beta}$. This is clear since every transform $Q_e$ reduces the length by a factor $\lambda^*$, and then we add a vector of length at most $\delta/\beta$. The length bound then is just a geometric series.


To prove the general case of the induction, we first show that it suffices to examine the problem in $2$ dimensions. So suppose that we have a deviation $r$ which satisfies that $\langle r, d^* \rangle \geq x^*$ for some constant $x^*$ to be determined later. We want to show that for any positive semi-definite $Q$ with all eigenvalues at most $\lambda^*$ and any $\vec{d}$ satisfying the length and inner product requirements above, that $\langle Q^\top r + d, d^* \rangle \geq x^*$.

It is clear that we only need to concern ourselves with the space of at most $3$ dimensions spanned by $d^*, r, Q^\top r$, since the added $d$ will add some fixed amount in the direction of $d^*$. We are concerned with how negative $\langle Q^\top r, d^* \rangle$ can be. We know that for any direction $z$, if $Q^\top r$ is in the direction $z$, the largest length it can have is $\|r\|_2 \cos{\theta}$ where $\theta$ is the angle between $z$ and $r$. Then in taking the inner product with $d^*$, we gain another factor $\phi$ where $\phi$ is the angle between $z$ and $r$. So if $Q^\top r$ is not in the same plane as $d^*$ and $r$, then since $\cos$ is increasing in the range $[0,\pi/2)$, we can replace $z$ with $z'$ which is the projection of $z$ into the plane of $d^*$ and $r$ then both $\theta$ and $\phi$ increase, and the resulting $\langle Q^\top r, d^* \rangle$ is more negative. Since we are concerned here with the worst case, it suffices here to consider only when $Q^\top r$ lies in the plane of $r$ and $d^*$, thus reducing the induction step to $2$ dimensions.


Now we prove the general step of the induction. We know that we can view this in 2 dimensions, so let us take $d^*$ to be the $x$-axis (recall that we defined it to be unit length, so we can do this without distorting lengths). Again, we are concerned with minimizing $\langle Q^\top r, d^* \rangle$. As we have seen, we can assume $Q$ is a scaled projection. So if it is onto a vector $z$ which forms an angle $\theta$ with $r$ and $\phi$ with the negative $x$-axis, then 

	\[ \langle Q^\top r, d^* \rangle 
				\geq \lambda^* \|r\|_2 \cos(\theta) \cos(\phi) 
				\geq \lambda^* \|r\|_2 \cos^2(\psi/2) \]

where $\psi$ is the angle between $r$ and the negative $x$-axis. Let $r = (x,y)$ now. Our inductive hypothesis is that $x \geq x^* \delta$. We also have $\phi = \tan^{-1}(y/x)$. Standard trigonometric manipulations (and careful choice of sign) give us that

	\[ \|r\|_2 \cos^2(\psi/2) \geq \frac12 (x - \sqrt{x^2 + y^2}) \geq \frac12 \left(x^* \delta - \frac{\delta}{1 - \lambda^*} \right) \]
\noindent
Our goal is to get
$\langle Q^\top r + d, d^* \rangle \geq \frac12 \left(x^* \delta - \frac{\delta}{1 - \lambda^*} \right) + x^* \delta$ to be $\geq x^* \delta$ 
Thus, to see what $x^*$ works, we solve for $x^*$ and get this is true if 
	\[ x^* \geq \frac{\beta - (\frac12 + \beta) \lambda^*}{(1 - \lambda^*)(1 - \frac12 \lambda^*)} \]

The expression on the right is positive when $\lambda^* < \frac{\beta}{\frac12 + \beta}$. In other words, when $\lambda^*$ satisfies this equation, an $x^*$ exists satisfying what we want.
This proves \eqref{eq:goal} for this generalized error model, using positive semi-definite matrices, and completes the proof of \Thm{directional}


\newpage
\appendix
\section{Proof of Claim \ref{claim:bounddelta}}\label{sec:omitted}
\bounddelta*
\begin{proof}
Note that $\Lambda(u) = 2 \sum_{i\ge 1} |L_i|\lambda^i$ where $L_i$ is the set of leaves at a distance $i$ from $u$. Since $\lambda < 1$, an $n$ leaf tree which maximizes $\Lambda(u)$ will make the tree as balanced in height as possible. (This can be proved by a ``swapping'' argument similar to the proof of optimality of Huffman trees.)  In particular, the maximizing tree has all leaves at distance $\floor{\log n}$ or
$\floor{\log n} + 1$. Therefore, $\Lambda(v) \le \frac{2}{\lambda}\cdot n\lambda^{\log n}= \frac{2}{\lambda}\cdot n^{1 - \log(1/\lambda)}$.
\end{proof}
\section{Proof of Fact \ref{lem:fact}}\label{sec:fact}
\lemfact*
\begin{proof}
Let $x$ be the vector with $||x||_1  = 1$ and $x\top {\mathbf 1} = 0$ such that $||P_1P_2|| = ||x^\top P_1P_2||_1$.
Note that $y^\top = x^\top P_1$ also satisfies $y^\top {\mathbf 1} = 0$ since $P_1$ is row stochastic. Therefore, $||y^\top P_2||_1 \leq ||y||_1\cdot ||P_2||$.
Also by definition, $||y||_1 = ||x^\top P_1||_1 \leq ||P_1||$.
\end{proof}

\section{Establishing the equation \eqref{eq:var}}\label{sec:eqvar}
Let us recall \eqref{eq:var}
\begin{align*}
\Var(Z_{v_i}|X_u = j) = \sum_{k\in \S} P_{jk}\Var(Z_{v_i}|X_v = k) + \frac{1}{2}\sum_{k\neq k' \in \S} P_{jk}P_{jk'}\Delta^2_{v_i}(k,k')
\end{align*}

We introduce some notational shorthand just to keep the exposition simple. We forgo the subscript on $Z_v$, let $V := \Var(Z| X_u = j)$, use ``$u = k$" to imply $X_u = k$, and use $\Exp^2[Z]$  to denote $(\Exp[Z])^2$.
Now, by definition,
$ V = \Exp[Z^2|u = j] - \Exp^2[Z|u = j] $.
\noindent
The first term evaluates to 
$$\Exp[Z^2|u=j]  = \sum_{k\in \S} P_{jk}\Exp[Z^2|v=k]$$ \notag 
\noindent
The second term evaluates to 
$$
\Exp^2[Z|u = j]  = \left( \sum_{k\in \S} P_{jk}\Exp[Z| v=k]\right)^2  = \sum_{k\in \S} P^2_{jk}\Exp^2[Z| v=k] + \sum_{k\neq k'\in \S} P_{jk}P_{jk'}\Exp[Z|v=k]\Exp[Z|v=k']$$
Observing $P^2_{jk} = P_{jk} - P_{jk}(1 - P_{jk})$, we get $V = $

$$\sum_{k\in \S}P_{jk}\left(\Exp[Z^2|v=k] - \Exp^2[Z|v=k]\right) + \sum_{k\in \S}P_{jk}(1-P_{jk})\Exp^2[Z|v=k] - \sum_{k\neq k'\in \S} P_{jk}P_{jk'}\Exp[Z|v=k]\Exp[Z|v=k']$$
\noindent
The first term above is  the first term in the RHS of \eqref{eq:var}. Furthermore, 
noting that $P_{jk}(1-P_{jk}) = \sum_{k\neq k' \in \S} P_{jk}P_{jk'}$ since $P_{jk}$'s sum up to $1$, we get that the second two terms is  
$$\frac{1}{2}\sum_{k\neq k'\in \S} P_{jk}P_{jk'}\left(\Exp^2[Z|v=k] + \Exp^2[Z|v=k'] - 2\Exp[Z|v=k]\Exp[Z|v=k']  \right) = \frac{1}{2}\sum_{k\neq k'\in \C} P_{jk}P_{jk'}\Delta^2_v(k,k')$$
\noindent
which establishes \eqref{eq:var}. 
\section{Proof of Lemma \ref{lem:models:negative}}\label{sec:negative}
 If the second eigenvalue is allowed to be close to 1, then there exists a sequence of transforms which causes the state-pair distribution at a leaf to be uniform, and thus indistinguishable from the independent case. In this section, we will focus on a 2-dimensional subspace of the error space containing $\vec{d}$. It is easy to check that as long as we choose a positive semi-definite transformation in this subspace of dimension 2, it is realizable in the full state-pair space of $s^2$ dimensions. We will let $d = (1,0)$ in this 2-dimensional subspace.

Now consider $k$ scaled projections $A_1, \ldots, A_k$ where

	\[ A_i = \lambda^* \left( \begin{array}{cc}
							\cos^2\left(\frac{i\pi}{k}\right) & \cos\left(\frac{i\pi}{k}\right) \sin\left(\frac{i\pi}{k}\right) \\
							\cos\left(\frac{i\pi}{k}\right) \sin\left(\frac{i\pi}{k}\right) & \sin^2\left(\frac{i\pi}{k}\right)
						\end{array} \right)
	\]

is a scaled projection onto a vector making an angle $i\pi/k$ with $\vec{d}$, the x-axis.

To make the deviation $\vec{r} = 0$ after the last transform ($\vec{r}^\top \,\mapsto\, \vec{r}^\top M + \vec{d}^\top$), we will first apply a large number of transforms where one of the eigenvectors is in the direction of $\vec{d}$, with a corresponding eigenvalue of $\lambda^*$. This will allow us to get our deviation $\vec{r}$ to be arbitrarily close to $\frac{1}{1 - \lambda^*} \vec{d}$. Then we will apply $A_1, \ldots, A_k$. We claim that if $\lambda^*$ is sufficiently large, this will be 0. Note that $A_k$ is a projection onto the $x$-axis, so we only have to examine the $x$ coordinates.

Before $A_1$, we have $\vec{r} = \frac{1}{1 - \lambda^*} \vec{d}$. After applying the transforms for $A_1, \ldots, A_k$, we have

	\[ r^\top A_1 \cdots A_k + d^\top (A_2 \cdots A_k + \ldots + A_k + \mathbf{I}_2) \]

where $\mathbf{I}_2$ is the 2-dimensional identity. We will examine the $x$ coordinates of these terms. For the first term, we see that each $A_i$ is a projection over an angle $\pi/k$ and includes a scaling $\lambda^*$, thus the $x$ coordinate of the first term is

	\[ \frac{-1}{1 - \lambda^*} (\lambda^* \cos(\pi/k) )^k \]

We will split the next part into two, as some of them will be negative and some will be positive. For $i = 1, \ldots, \lceil k/2 \rceil -2$, these will contribute negatively the amount

	\[ -(\lambda^* \cos(\pi/k))^{n - i - 1} \cos((i+1)\pi/k) \]

For $i = \lceil k/2 \rceil - 1, \ldots, k - 1$, this will contribute positively the amount

	\[ (\lambda^* \cos(\pi/k))^{n - i - 1} \cos((n - i - 1) \pi/k) \]

Finally, $\vec{d}^\top \mathbf{I}_2$ contributes 1. In total, this gives

	\begin{eqnarray*}
		&&- \left(\frac{1}{1 - \lambda^*} (\lambda^* \cos(\pi/k) )^k 
					+ \sum_{i=1}^{\lceil r/2 \rceil - 2} 
							(\lambda^* \cos(\pi/k))^{n - i - 1} \cos((i+1)\pi/k) \right) \\
		&&+ \left(1 
					+ \sum_{i=\lceil r/2 \rceil - 1}^{k-1} 
							(\lambda^* \cos(\pi/k))^{n - i - 1} \cos((n - i - 1) \pi/k) \right)
	\end{eqnarray*}

Finally, for a fixed $k$ we can solve for this to be 0 to get an upper bound on allowable $\lambda^*$. For $k = 9$, this gives $\lambda^* < 0.832$. 


\begin{thebibliography}{99}

\bibitem{Ca78} {\sc J.A. Cavender.} \newblock Taxonomy with Confidence. \newblock {\em Math. Biosci,} {\bf 40}: 271--80 (1978).

\bibitem{Chang} {\sc J. T. Chang.}
\newblock Full Reconstruction of Markov Models on Evolutionary Trees: Identifiability and Consistency.
\newblock {\em Mathematical Biosciences}, 137, 51--73, 1996.

\bibitem{DaMoRo06} {\sc C. Daskalakis, E. Mossel, and S. Roch.} \newblock Optimal phylogenetic reconstruction. \newblock {\em Proc. 38th ACM STOC,} 159--166 (2006).

\bibitem{Erdos1}{\sc P.L. Erd\"os, M. Steel, L. Szekely and T. Warnow.}
\newblock A few logs suffice to build (almost) all trees (I).
\newblock {\em Random Structure and Algorithms,} 14, 153--184, 1997.

\bibitem{Erdos2}{\sc P.L. Erd\"os, M. Steel, L. Szekely and T. Warnow.}
\newblock A few logs suffice to build (almost) all trees (II).
\newblock {\em Theoretical Computer Science}, 221 (1--2), 77--118, 1999.


\bibitem{Fa73} {\sc J.S. Farris.} \newblock A probability model for inferring evolutionary trees. \newblock {\em Syst. Zool.} {\bf 22}:250--56 (1973).

\bibitem{FarachKannan} {\sc M. Farach and S. Kannan.}
\newblock Efficient algorithms for inverting evolution.
\newblock {\em Proc. 28th ACM STOC,} 1996.

\bibitem{Fe81} {\sc J. Felsenstein.} \newblock Evolutionary trees from DNA sequences: a maximum likelihood approach.
\newblock {\em J. Mol. Evol.} {\bf 17}:368--76 (1981).

\bibitem{Fe04} {\sc J. Felsenstein.} \newblock {\em Inferring Phylogenies.} \newblock Sinauer, New York, 2004.


\bibitem{HuCr97} {\sc J. Huelsenbeck and K. Crandall.} \newblock Phylogeny estimation and hypothesis testing using maximum likelihood.
\newblock {\em Annu. Rev. Ecol. Syst.} {\bf 28}:437--66 (1997).

\bibitem{JC69}{\sc T. H. Jukes, and C. R. Cantor}
\newblock Evolution of Protein Molecules.
\newblock {\em New York: Academic Press}, 1969.

\bibitem{Kimura80}{\sc M. Kimura}
\newblock A simple method for estimating evolutionary rates of base substitutions through comparative studies of nucleotide sequence.
\newblock {\em J. Mol. Evol.} 16(2): 111--120, 1980.

\bibitem{peres-book} {\sc D. A. Levin, Y. Peres, and E. L. Wilmer}
\newblock Markov Chains and Mixing Times.
\newblock {\em American Mathematical Society}, ISBN-10: 0-8218-4739-2, 2008.

\bibitem{Ma90} {\sc W. Maddison.} \newblock A Method for testing the correlated evolution of two binary characters: are gains or losses concentrated on certain branches of a phylogenetic tree? \newblock {\em Evolution,}
{\bf 44}(3), 539--557, 1990.

\bibitem{MJH+13}{\sc F. Morcos, B. Jana, T. Hwa, and J. Onuchic}
\newblock Co-evolutionary signals across protein lineages help capture multiple protein conformations
\newblock {\em PNAS} 110:20533--20538 (2013).

\bibitem{MPL+11}{\sc F. Morcos, A. Pagnini, B. Lunt, A. Bertolino, D. Marks, C. Sander, R. Zecchina, J.N. Onuchic, T. Hwa and M. Weigt}
\newblock{Direct-coupling analysis of residue co-evolution captures native contacts across many protein families}
\newblock {\em PNAS} 108: 1293--1301 (2012).

\bibitem{Mo04} {\sc E. Mossel.} \newblock Phase transitions in phylogeny. \newblock {\em Trans. Amer. Math. Soc.} {\bf 356}:6 2379--2404 (electronic) 2004.

\bibitem{MosselRoch} {\sc E. Mossel and S. Roch.}
\newblock Learning Nonsingular Phylogenies and Hidden Markov Models.
\newblock {\em Proc. of 37th ACM STOC}, 2005.

\bibitem{Ne71} {\sc J. Neyman.} \newblock Molecular studies of evolution: a source of novel statistical problems. \newblock In {\em Statistical decision theory and related topics,} S.S. Gupta
and J. Yackel (eds.) 1--27 (1971).

\bibitem{SS03} {\sc C. Semple and M. Steel.} \newblock {\em Phylogenetics,}. \newblock Oxford Lecture Series in Mathematics and its Applications. {\bf 24}. 

%



\end{thebibliography}
\end{document}